\documentstyle[aps,prl]{revtex}

\newcommand{\kk}{{\bf k}}
\newcommand{\pp}{{\bf p}}
\newcommand{\qq}{{\bf q}}
\newcommand{\xx}{{\bf x}}
\newcommand{\yy}{{\bf y}}

\newcommand{\BE}{\begin{equation}}
\newcommand{\EE}{\end{equation}}
\newcommand{\BA}{\begin{eqnarray}}
\newcommand{\EA}{\end{eqnarray}}

\begin{document}
\draft

\title{Condensation of gauge interacting mass-less fermions}

\author{Fabio Siringo}
\address{Dipartimento di Fisica e Astronomia, 
Universit\`a di Catania,\\
INFN Sez. di Catania and INFM UdR di Catania,\\
Via S.Sofia 64, Catania, Italy}
\date{\today}

\maketitle

\begin{abstract}
A single mass-less fermionic field with an abelian U(1) gauge interaction
(electrodynamics of a mass-less Dirac fermion) is studied by a variational
method. Even without the insertion of any extra interaction the vacuum is shown
to be unstable towards a particle-antiparticle condensate. The single particle
excitations do acquire a mass and behave as massive Fermi particles. 
An explicit low-energy gap equation has been derived and numerically
solved. Some consequences of condensation and mass generation 
are discussed in the framework of the standard model.
\end{abstract}


\section{Introduction}

One of the most outstanding problems in the Standard Model (SM)
of strong and electroweak interactions is the very nature of the
Higgs boson. That heavy scalar particle has not been found yet, but
is required by the theory in order to give a mass to all the
fermions through the standard symmetry breaking mechanism.
The nature of the scalar field is not established and
several composite models have been proposed.  Top-quark
condensation\cite{top}
is one of the most interesting and economical
mechanisms as it does not require the existence of any new particle.
The basic idea goes back to Nambu and Jona-Lasinio's\cite{nambu}
work on fermion
condensation which appeared in 1961, just four years later than
the celebrated paper on superconductivity by Bardeen, Cooper and
Schrieffer (BCS)\cite{bcs}.
The formal analogy between the single particle excitation
energies of a superconductor $\epsilon_k=\sqrt{\kk^2+\Delta^2}$ and
the relativistic mass-shell equation (vacuum excitation energies) for
a massive particle is striking and would suggest that
the mass is just the gap $\Delta$ which opens in some
condensation mechanism.
However, all Top-quark condensation models require the existence of an
{\it ad hoc} four-fermion interaction which must be added to the standard
lagrangian.
Then the SM is recovered as a low energy effective theory with the
scalar Higgs field describing the condensate.

In this paper we discuss the problem of fermion condensation 
without adding any four-fermion attraction. In fact by a variational method
we show that the
standard gauge interactions may give rise to condensation of the
original massless fermions. We address the problem by use of a simplified
toy model consisting of a single massless fermionic field with an
abelian $U(1)$ gauge interaction
(electrodynamics of a massless Dirac fermion).
In this framework condensation may have several important consequences:
first of all the SM consists of gauge interacting massless fermions, and
their {\it spontaneous} condensation could spoil the standard symmetry
breaking and mass generation mechanisms which are supposed to be due to
the interaction with the scalar field. Moreover, a {\it spontaneous}
condensation could replace the standard mechanism without having to
insert any extra field or interaction.

Again the idea that gauge interactions could give rise to condensation
of massless fermions is not new, and comes from condensed matter.
Ten years after the BCS paper\cite{bcs}, Jerome, Rice and Kohn\cite{rice}
predicted the
existence of an {\it excitonic insulator} in gapless semiconductors
as a consequence of the condensation of electron-hole pairs bound by
the coulomb interaction. Since the pair (the exciton) does not carry
any charge, the $U(1)$ symmetry is unbroken, and condensation does not
give rise to any superconductive property. However a gap opens in the
spectrum of single particle excitation energies, and the quasi-particles
do acquire a mass (the system is predicted to be an insulator).

For a generic charged massless fermionic field we predict the occurrence of
a spontaneous condensation for any weak long range coupling: as single particle
excitation energies are gapless in the original vacuum,
any weak long range attraction would drive the condensation of
particle-antiparticle pairs.
In the new vacuum the single particle excitation energies (the physical
particles) acquire a finite mass which obviously depends on the only energy
scale of the problem: the energy cut-off required in order to regularize the
theory.

In our toy model the mass is very small compared to fermionic
masses, unless the cut-off is allowed to take huge values. Thus below
the grand unified scale ($\sim 10^{16}$ GeV) spontaneous condensation
should not spoil the SM symmetry breaking mechanism. However
a mass as large as the electron rest mass can be recovered by a
cut-off approaching the Landau singularity point.
In the light of these findings,
the full non-abelian theory should be studied in order to achieve a
quantitative estimate of the effect which could be relevant
in the context of top-quark condensation. On the other hand, a small
mass could also arise for neutrinos, as a consequence of weak interactions.
We notice that in general this spontaneous condensation would break chiral
invariance while leaving the $U(1)$ symmetry unbroken. 
Such open problems, while motivating our work, go beyond
the aims of the present paper which only deals with the existence of
spontaneous condensation and is organized as follows: in the next 
section the toy model is defined and the variational method is described; in 
section III a gap equation is obtained and details of the derivation 
are presented; section IV contains a full discussion of the analytical results and
comments on their relevance; a low-energy gap equation is derived in
section V where a more formal proof  is given of the existence of a 
non-trivial solution. In that section the problem of mass generation is discussed
by numerical solution of the low-energy gap equation.

\section{The model}

Let us consider the following toy-model lagrangian:
\BE
{\cal{L}}=
-\bar\Psi\left(\gamma^\mu \partial_\mu +ie\gamma^\mu A_\mu\right) 
\Psi -{1\over 4} F_{\mu\nu}F^{\mu\nu}
\EE
It containes a Dirac mass-less $S={1\over 2}$ fermionic field
$\Psi$ with a $U(1)$ (e.m.) gauge interaction field $A_\mu$, and
$F_{\mu\nu}=\partial_\mu A_\nu-\partial_\nu A_\mu$.
The electric current is
\BE
J^\mu={{\partial\cal{L}}\over{\partial A_\mu}}=
-ie\bar\Psi\gamma^\mu\Psi
\EE
In Coulomb gauge the hamiltonian reads
\BE
H=H_0+V_c+V_i
\label{H}
\EE
where $H_0$ is the free particle hamiltonian
\BE
H_0=\int d^3 p\sum_\sigma \vert \pp\vert
a^\dagger(\pp,\sigma)a(\pp,\sigma)+
\int d^3p\sum_\sigma \vert\pp\vert
\left[\alpha^\dagger (\pp,\sigma)\alpha(\pp,\sigma)
+\beta^\dagger (\pp,\sigma)\beta(\pp,\sigma)\right]
\EE
(here $a,a^\dagger$ are photon annihilation,creation operators,
while $\alpha,\alpha^\dagger$ and $\beta,\beta^\dagger$ are
particle and anti-particle operators for the mass-less fermions),
$V_c$ is the Coulomb interaction
\BE
V_c={1\over 2}\int d^3x\int d^3 y{{J^0(x) J^0 (y)}\over
{4\pi\vert \xx-\yy\vert}}
\EE
and $V_i$ is the interaction term
\BE
V_i=-\int d^3 x {\bf J} (x) {\bf A} (x).
\EE

Let us take as a trial vacuum the BCS-like vacuum discussed by
Nambu and Jona-Lasinio\cite{nambu}
\BE
\vert\Phi\rangle=\prod_{\kk,\sigma}
\left[h_k+g_k\alpha^\dagger(\kk,\sigma)\beta^\dagger(-\kk,\sigma)
\right]\vert 0\rangle
\EE
where $\vert 0\rangle$ is the vacuum annihilated by $a,\alpha,\beta$.
For $g_k=0$ this trial vacuum contains the trivial 
vacuum $\vert 0\rangle$. For $g_k\not=0$ the trial vacuum is 
characterized by pair correlation: each pair has zero charge, 
zero momentum and zero spin ($\sigma$ denotes helicity). Thus, at
variance with superconductivity, we expect e.m. $U(1)$ gauge 
invariance to be unbroken (Eventually $SU(2)_L$ could be broken
since the pair carries two units of chirality).

Next let us evaluate the ground state energy 
\BE
E=\langle\Phi\vert H\vert \Phi\rangle
\EE
The average value of $V_i$ vanishes (it is linear in
$a$ and $a^\dagger$), and we only need
\BE
E=\langle\Phi\vert H_0\vert \Phi\rangle
+\langle\Phi\vert V_c\vert \Phi\rangle
\label{energy}
\EE
Actually $V_c$ is a four-fermion correlation interaction:
\BE
V_c={1\over 2}\int {{d^3 q}\over{(2\pi)^3}}
{{J^0(\qq)J^0(-\qq)}\over{\qq^2}}
\label{coulomb}
\EE
where $J^0(\qq)$ is the Fourier transform
\BE
J^0(\qq)=\int d^3 x e^{-i\qq\xx} 
\left[-e\Psi^\dagger (x)\Psi(x)\right]
\label{fourier}
\EE
and the fermion field $\Psi(x)$ is given as usual by
\BE
\Psi(x)={1\over {(2\pi)^{3/2}}}\int d^3 p\sum_\sigma
\left[e^{i\pp\xx} u(\pp,\sigma)\alpha(\pp,\sigma)
+e^{-i\pp\xx} v(\pp,\sigma)\beta^\dagger (\pp,\sigma)\right].
\label{field}
\EE
Insertion of Eq.(\ref{field}) and (\ref{fourier}) in 
Eq.(\ref{coulomb}) yields a four-fermion interaction.
The average energy $E$ in Eq.(\ref{energy}) may be evaluated
by use of the properties of the spinors $u$ and $v$ 
(spin sums and traces).

The coefficients $g_k$ and $h_k$ are then regarded as variational
parameters, with the normalization constraint 
\BE
\vert g_k\vert^2+\vert h_k\vert^2=1
\label{norma}
\EE
and the true vacuum (ground state) is recovered by differentiating
Eq.(\ref{energy}).

\section{The gap equation}

The explicit evaluation of Eq.(\ref{energy}) is straightforward.
The average value of $V_c$ reads
\BA
\langle\Phi\vert V_c\vert \Phi\rangle&=&
e^2\Omega\sum\int {{d^3pd^3p^\prime d^3 q}\over {(2\pi)^3 \qq^2}}\times\cr
&\times&\langle\Phi\vert 
\eta_i^\dagger(\pp-\qq,\sigma)
\eta_j(\pp,\tau)
\eta_k^\dagger(\pp^\prime+\qq,\sigma^\prime)
\eta_l(\pp^\prime,\tau^\prime)
\vert\Phi\rangle\times\cr
&&\qquad\qquad\qquad\qquad\times S_i^\dagger(\pp-\qq,\sigma)
S_j(\pp,\tau)
S_k^\dagger(\pp^\prime+\qq,\sigma^\prime)
S_l(\pp^\prime,\tau^\prime)
\label{coulomb2}
\EA
where $\Omega$ is the space volume and the sum runs over helicities $\sigma$,$\sigma^\prime$,$\tau$,$\tau^\prime$
and over $ijkl=1,2$, with the notation
\BA
\eta_1^\dagger(\pp,\sigma)&=&\alpha^\dagger(\pp,\sigma)\cr
\eta_2^\dagger(\pp,\sigma)&=&\beta(-\pp,\sigma)\cr
S_1(\pp,\sigma)&=&u(\pp,\sigma)\cr
S_2(\pp,\sigma)&=&v(-\pp,\sigma).
\label{notation}
\EA
This average vanishes unless any $\eta^\dagger(\pp,\sigma)$ is joined by a
$\eta(\pp,\sigma)$ with the same arguments: as the trial state $\vert\Phi\rangle$ 
only contains pair correlations the action of any creation operator must be followed
by the annihilation of the same particle or by the creation of the paired antiparticle.
We have three cases: i) $\qq=0$, $\sigma^\prime=\tau^\prime$, $\sigma=\tau$
(with $\pp\not=\pp^\prime$ or $\sigma\not=\sigma^\prime$); ii) $\qq=0$,
$\pp=\pp^\prime$, $\sigma=\sigma^\prime=\tau=\tau^\prime$;
iii)$\pp-\pp^\prime=\qq$, $\sigma=\tau^\prime$, $\tau=\sigma^\prime$
(with $\qq\not=0$ or $\sigma\not=\sigma^\prime$).
Both cases i) and ii) only contribute constant terms (in the sense that they do not depend
on $g_k$) and may be dropped as they give the same energy contribution in the
trivial vacuum $\vert 0\rangle$. For the case iii) the sum over helicities yields
\BE
\sum_{\sigma\tau}
S_i^\dagger(\pp^\prime,\sigma)
S_j(\pp,\tau)
S_k^\dagger(\pp,\tau)
S_l(\pp^\prime,\sigma)
=Tr\left[A_{jk}(\pp)A_{li}(\pp^\prime)\right]
\label{spinsum}
\EE
where for each pair $(ij)$ the matrix $A_{ij}$ is
\BE
A_{ij}(\pp)=\sum_\sigma S_i(\pp,\sigma) S_j^\dagger (\pp, \sigma).
\EE
The non-vanishing contributions are
\BA
Tr\left[A_{11}(\pp)A_{11}(\pp^\prime)\right]&=&1+
{{\pp\cdot\pp^\prime}\over{\vert\pp\vert\vert\pp^\prime\vert}}
\cr
Tr\left[A_{11}(\pp)A_{22}(\pp^\prime)\right]&=&1
-{{\pp\cdot\pp^\prime}\over{\vert\pp\vert\vert\pp^\prime\vert}}
\cr
Tr\left[A_{12}(\pp)A_{12}(\pp^\prime)\right]=
Tr\left[A_{21}(\pp)A_{21}(\pp^\prime)\right]&=&1-
{{\pp\cdot\pp^\prime}\over{\vert\pp\vert\vert\pp^\prime\vert}}
\cr
Tr\left[A_{12}(\pp)A_{21}(\pp^\prime)\right]=
Tr\left[A_{21}(\pp)A_{12}(\pp^\prime)\right]&=&1+
{{\pp\cdot\pp^\prime}\over{\vert\pp\vert\vert\pp^\prime\vert}}.
\label{tacce}
\EA
Inserting the coefficients (\ref{spinsum}) for the respective averages
$\langle \eta_i^\dagger\eta_j\eta_k^\dagger\eta_l\rangle$ we obtain
\BE
\langle\Phi\vert V_c\vert \Phi\rangle=E_{EXC}+E_{BCS}
\EE
where
\BE
E_{EXC}=2e^2\Omega\int{{d^3p d^3 k}\over{(2\pi)^6}}
{{\vert g_p\vert^2\vert h_{k}\vert^2 \pp\cdot\kk}\over
{\vert\pp-\kk\vert^2 \vert\pp\vert\vert\kk\vert}}
\label{EXC}
\EE
\BE
E_{BCS}=-2e^2\Omega \int{{d^3p d^3k}\over{(2\pi)^6}}
{{h_p g_p h_k^*g_k^*}\over
{\vert\pp-\kk\vert^2}}
\label{BCS}
\EE
The average of $H_0$ is trivial
\BE
\langle\Phi\vert H_0\vert \Phi\rangle=E_0=4\Omega\int{{d^3p}\over{(2\pi)^3}}
\vert g_p\vert^2\vert \pp\vert
\EE
and the total energy  reads
\BE
E=E_0+E_{BCS}+E_{EXC}
\label{Etotal}
\EE

While $E_{BCS}$ is the usual BCS pairing energy, the term $E_{EXC}$ survives from a partial compensation
of  the exchange energies in the particle-antiparticle condensate. As the effects of this exchange term can be
dealt with by standard perturbative renormalization of parameters, we neglect this term in order to simplify 
the gap equation. In  the appendix we show that the inclusion of the exchange term would give rise to a
charge renormalization which is equivalent to the standard perturbative renormalization up to first order.
In the following discussion we assume that both charge and mass are the physical renormalized values in order
to incorporate the effects of the exchange energy and of the other interactions which have been neglected in 
this simple toy model. Conversely the term $E_{BCS}$ cannot be dealt with by standard perturbation theory
as it makes the trivial vacuum unstable. As usual we attempt a variational estimate of  the best $g_p$ value by
differentiating the total energy $E=E_0+E_{BCS}$.

According to the normalization condition (\ref{norma})  we denote
\BA
g_p&=&\sin\theta_p\cr
h_p&=&\cos\theta_p
\label{theta}
\EA
and differentiate the energy
\BE
{1\over\Omega}{{\delta E}\over{\delta \theta_p}}=
4\vert \pp\vert\sin 2\theta_p-2e^2\cos 2\theta_p\int
{{d^3k}\over{(2\pi)^3}}{{\sin 2\theta_k}\over{\vert\pp-\kk\vert^2}}=0.
\label{derivative}
\EE
This can be written as a standard gap equation
\BE
\Delta_p={{e^2}\over 2}\int
{{d^3k}\over{(2\pi)^3}}
{1\over{\vert\pp-\kk\vert^2}} 
{{\Delta_k}\over{\sqrt{\kk^2+\Delta_k^2}}}
\label{gap}
\EE
where the gap function $\Delta_k$ is defined according to
\BE
{{\Delta_p}\over{\vert\pp\vert}}={{\sin 2\theta_p}\over{\cos 2\theta_p}}=
{{e^2}\over{2\vert\pp\vert}}
\int{{d^3k}\over{(2\pi)^3}}{{\sin 2\theta_k}\over{\vert\pp-\kk\vert^2}}.
\label{gapfun}
\EE
Eq.(\ref{gap}) is the gap equation we would have expected from the beginning.
The trivial vacuum $\vert 0\rangle$ is given by the solution $\Delta_k=0$ which is
not the ground state as an other unconventional solution can be always found for
any strength of the coupling constant.

\section{Condensation and fermion masses}
The  non trivial solution of the gap equation (\ref{gap}) describes a particle-antiparticle
condensate. The single particle excitations are the ``physical'' particles and are characterized
by the energy spectrum
\BE
\varepsilon_k=\sqrt{\kk^2+\Delta_k}.
\label{spectrum}
\EE
Thus the single particle excitations behave like massive fermions with a
mass $M=\Delta_0$. The IR behaviour of Eq.(\ref{gap}) requires the
existence of a non vanishing mass $M$. In fact at low energy,
replacing $\Delta_k\approx M$ the gap equation reads
\BE
{2\over {e^2}}=\int{{d^3k}\over{(2\pi)^3}}
{1\over {\kk^2\sqrt{\kk^2+M^2}}}
\label{masseq}
\EE
and the logarithmic divergence for $\kk\to 0$ ensures that the
mass M is not vanishing. The exact value depends on the 
high energy behaviour of the function $\Delta_k$: the UV convergence of
the integrals requires that $\Delta_k$ should be vanishing at large energies.
We may estimate $M$ by the ansatz $\Delta_k=0$ for $\vert\kk\vert>\Lambda$
which is equivalent to the insertion of a cut-off $\Lambda$ in Eq.(\ref{masseq})
thus obtaining
\BE
M\approx 2\Lambda e^{ -{{4\pi^2}/{e^2}}}.
\label{mass}
\EE
Thus the variational method shows that a non-vanishing mass $M$ is always present for
any weak coupling $e^2$. As for superconductivity this result cannot be obtained by any perturbative 
expansion in $e^2$ starting from the trivial vacuum $\vert 0\rangle$. 
Of course the mass $M$  depends on the unique 
length scale of the model which is the energy cut-off  $\Lambda$. Moreover the mass $M$ is very
small unless the cut-off $\Lambda$ is supposed to be really huge. The scale $\Lambda$ would represent
an intrinsic limit of the simple electrodynamics of a charged fermion. It is remarkable that, assuming for
$M$ the phenomenological value of the electron mass, the cut-off  $\Lambda$ reaches the large value
\BE
\Lambda\approx {M\over 2}e^{ {{4\pi^2}/{e^2}}}
\EE
close  to the Landau singularity
\BE
\Lambda_{Landau}=Me^{ {{6\pi^2}/{e^2}}},
\EE
but still smaller.

We notice that for any weak coupling, the occurrence of particle-antiparticle condensation and the 
opening of a gap are the natural  consequence of  two important aspects:
i) the vanishing of mass in the trivial vacuum; ii) the long range behaviour of the Coulomb interaction.

The trivial vacuum for mass-less fermions is unstable because the creation of a particle-antiparticle pair
does not require any energy, while the particle and the antiparticle 
attract each other: thus the creation energy may
become negative. This is known to be the case for semiconductors when the gap is vanishing.
However, at variance with condensed matter, the vanishing of the gap is not by itself a sufficient condition for
determining the vacuum instability. In superconductors the integration over $\kk$ is carried across 
the Fermi level, where $\vert\kk\vert=k_F$ is a finite Fermi vector, and the measure only gives a simple $dk$
factor
which is not enough for the IR convergence. The opening of the gap is then necessary as otherwise in the
gap equation the integral would diverge logarithmically.
Conversely here the integration reaches the $\kk=0$ point and the measure gives a $\kk^2 dk$ term which
would be enough for the IR convergence of the integral in Eq.(\ref{gap}) were it not for the extra 
IR divergence of  the Coulomb interaction $e^2/\kk^2$. Actually it is well known that for any short-range
interaction condensation only takes place if the coupling is strong enough\cite{nambu}. The diverging
$1/\kk^2$ behaviour of the Coulomb interaction here cancels the extra measure factor $\kk^2$ thus
restoring the IR logarithmic divergence in the gap equation:
the long range behaviour of gauge interactions is a key factor for the condensation of mass-less
fermions.

\section{Mass generation and low-energy gap equation}

In the light of the present study we believe it correct to say that fermion condensation does not spoil the
SM mechanism of symmetry breaking, as far as fermions do acquire a mass by interaction with
the Higgs field or by some other effect. However fermion condensation could itself be a candidate to
such symmetry breaking description with the condensate playing the role of the scalar field.
In that framework  any further discussion on mass generation would require
a numerical solution of the gap equation Eq.(\ref{gap}), at least in the low-energy domain. Unfortunately
the huge ratio $\Lambda/M$ rules out a direct numerical evaluation, as the gap turns out to be too small
for any viable cut-off $\Lambda$. In fact any numerical attempt to solve Eq.(\ref{gap}) would question
the existence of any non-trivial solution. However a non-vanishing numerical solution is easily found by
iteration as long as the coupling parameter $e^2$ is taken large enough to keep the ratio $\Lambda/M$
at a numerically tractable value according to Eq.(\ref{mass}). From the existence of a non-trivial solution at
large coupling we may prove that the gap must remain non-vanishing for any weak coupling. In fact, should
the gap vanish at a critical coupling  $e^2=e_c^2$, we could expand the gap equation Eq.(\ref{gap})
in powers of $\Delta_k$ at that critical point. The first order approximation is easily obtained by inserting
$\Delta_k=0$ in the square root, and holds for a vanishing gap. Thus at the critical point the gap equation
can be replaced by the eigenvalue problem of a linear integral operator. The existence of a non-trivial solution
for $e^2>e_c^2$ is equivalent to say that $1/e^2_c$ is the larger eigenvalue of the linear integral operator. 
But the integral operator is not bounded as the kernel diverges for $p\to 0$, $k\to 0$, and thus $e^2_c= 0$.
That means the non-trivial solution becomes very small for a weak coupling but does not vanish as long as the
coupling is $e^2>0$.

A more tractable low-energy gap equation can be derived for the realistic weak coupling limit
$e^2/(4\pi)=\alpha\approx 1/137$ of quantum electrodynamics. 
Let us consider the arbitrary intermediate scale $\mu$, assuming $\Delta_p\ll\mu$ for any $p$, but
$\mu\ll\Lambda$. By integrating over angles Eq.(\ref{mass})
reads
\BE
\Delta_p={\alpha\over{2\pi}}\int_0^\mu dk~ {k\over p} \ln
\left\vert{{p+k}\over{p-k}}\right\vert
{{\Delta_k}\over{\sqrt{k^2+\Delta_k^2}}}
+{\alpha\over{2\pi}}\int_\mu^\Lambda dk~ {k\over p} \ln
\left\vert{{p+k}\over{p-k}}\right\vert
{{\Delta_k}\over{\sqrt{k^2+\Delta_k^2}}}
\EE
In the low energy domain $p\ll\mu$ the second integral may be approximated by
\BE
{\alpha\over{2\pi}}\int_\mu^\Lambda dk~ {k\over p} \ln
\left\vert{{p+k}\over{p-k}}\right\vert
{{\Delta_k}\over{\sqrt{k^2+\Delta_k^2}}}
={\alpha\over{\pi}}\int_\mu^\Lambda dk~ {{\Delta_k}\over k}
+{\cal O} \left(p^2/\mu^2\right)
\EE
since $p\ll\mu<k$ and $\Delta_k\ll \mu<k$ so that $\sqrt{\Delta_k^2+k^2}\approx k$.
For small $p$ we may drop the dependence on $p$ and write
\BE
\Delta_p=M_\mu+{\alpha\over{2\pi}}\int_0^\mu dk~ {k\over p} \ln
\left\vert{{p+k}\over{p-k}}\right\vert
{{\Delta_k}\over{\sqrt{k^2+\Delta_k^2}}}
\label{gap_low}
\EE
where any dependence on $\Lambda$ is now in the renormalized mass $M_\mu$
which is defined at the scale $\mu$ as
\BE
M_\mu={\alpha\over{\pi}}\int_\mu^\Lambda dk~ {{\Delta_k}\over k}.
\EE
Eq.(\ref{gap_low}) is an approximate low-energy gap equation for $\Delta_k$ in the
range $k<\mu$. The high energy behaviour of $\Delta_k$ would only be required in order to
evaluate the renormalized mass constant $M_\mu$. However, by a proper choice of $\Lambda$,
$M_\mu$ can take any chosen value and can be regarded as a free parameter incorporating any
dependence on the cut-off. Moreover $M_\mu$ must be close to the phenomenological mass as
the integral in Eq.(\ref{gap_low}) yields a very small contribution. The low-energy gap equation
Eq.(\ref{gap_low}) always has  a solution which can be easily evaluated by  numerical iteration.
The numerical solution is shown in Fig.1 for the phenomenological coupling $\alpha=1/137$ and
for two different scales: a small scale $\mu=1.5 M_\mu$,
and a large scale $\mu=220 M_\mu$. For small $p$ the gap $\Delta_p$ is almost constant, with a
decrease of less than 0.15\% in the range   $0<p<2M_\mu$ of the large scale solution. Thus we may
regard the $p\to 0$ limit of $\Delta_p$ as the ``physical'' mass $M=\Delta_0$.

A perturbative iterative solution in powers of  the coupling $\alpha$ may be written for the
low-energy gap equation by inserting
$\Delta_p=M_\mu+{\cal O} (\alpha)$ in the right hand side of Eq.(\ref{gap_low}), yielding
in the limit $p\to 0$
\BE
M\approx M_\mu+{\alpha\over\pi}M_\mu\ln\left(\mu\over{M_\mu}\right)+
{\cal O}\left(\alpha^2\right)
\EE
This first order mass shift turns out to be just $2/3$ times the standard perturbative
self-energy contribution at the energy scale $\mu$.

While  fermion condensation would be the simplest mass generation mechanism, it would
also provide the existence of a condensate playing the role of the Higgs field. 
Moreover, as pairs carry a unit of helicity, any heli-magnetic solution would give a simple 
picture of left-right symmetry breaking. 
Quite interesting, as weak interactions are long ranged before symmetry breaking, even neutrinos
could undergo condensation and would acquire a very small mass.
All these outstanding problems call for
a detailed study of the full non-abelian gauge group which goes beyond the aim of the present
paper. 

\vskip 20pt

\appendix
\section{Exchange energy and charge renormalization}

Inclusion of the exchange term $E_{EXC}$ Eq. (\ref{EXC}) in the total energy would
only give rise to a more complicated gap equation which can be cast again in the shape of
Eqs.(\ref{gap}) and (\ref{masseq}) by charge renormalization.
Insertion of Eq.(\ref{theta}) and differentiation yield the following gap equation
\BE
\Delta_p={{e^2}\over{ 2(1+I_p)}}\int
{{d^3k}\over{(2\pi)^3}}
{1\over{\vert\pp-\kk\vert^2}} 
{{\Delta_k}\over{\sqrt{\kk^2+\Delta_k^2}}}
\label{gap2}
\EE
where the gap function $\Delta_k$ is defined according to
\BE
{{\Delta_p}\over{\vert\pp\vert}}={{\sin 2\theta_p}\over{\cos 2\theta_p}}=
{{e^2}\over{2\vert\pp\vert (1+I_p)}}
\int{{d^3k}\over{(2\pi)^3}}{{\sin 2\theta_k}\over{\vert\pp-\kk\vert^2}}.
\label{gapfun2}
\EE
and the integral $I_p$ is
\BE
I_p={{e^2}\over{2\vert\pp\vert}}
\int{{d^3k}\over{(2\pi)^3}}
{{(\kk\cdot\pp) \cos 2\theta_k  }\over
{\vert\pp\vert\vert\kk\vert\vert\pp-\kk\vert^2}}.
\label{Ip}
\EE
For small energies
\BE
I_p=e^2 A+{\cal O}(p)
\label{Ip2}
\EE
where the constant $A$ is given by the integral
\BE
A={{1}\over{ 6\pi^2}}\int_0^\Lambda
{{dk}\over{\sqrt{k^2+\Delta_k^2}}}.
\label{A}
\EE
which is regularized by a cut-off $\Lambda$.
Thus for $M=\Delta_0$ Eq.(\ref{masseq}) is recovered as
\BE
{2\over {Z_3e^2}}=\int{{d^3k}\over{(2\pi)^3}}
{1\over {\kk^2\sqrt{\kk^2+M^2}}}
\label{masseq2}
\EE
where the renormalization constant $Z_3$ reads
\BE
Z_3={1\over{1+e^2A}}
\EE
and up to  order $e^2$, inserting Eq.(\ref{A}), we obtain
\BE
Z_3\approx 1-{{e^2}\over {6\pi^2}}\ln(\Lambda/M)
\label{Z3}
\EE
which is exactly the standard 
first order perturbative result for $Z_3$.
As shown by Eq.(\ref{masseq2}) all the discussions on Eq.(\ref{masseq})  
still hold provided that the renormalized coupling $Z_3 e^2$ is substituted
for the bare coupling $e^2$.

\begin{figure}
\caption{Numerical solution of the low-energy gap equation Eq.(\ref{gap_low})
for $\alpha=1/137$. 
The gap $\Delta_p$ is reported as a function of $p$ in units of the mass constant $M_\mu$,
and for two different energy scales $\mu=1.5 M_\mu$ (solid line) and $\mu=220 M_\mu$
(dashed line).}
\end{figure}


\begin{thebibliography} {99}
  \bibitem{top} For a review see, for instance, G. Cvetic,
  Rev. Mod. Phys. {\bf 71}, 513 (1999).
  \bibitem{nambu} Y.Nambu and G.Jona-Lasinio,
  Phys.Rev. {\bf 122}, 345 (1961).
  \bibitem{bcs} J.Bardeen, L.N.Cooper, J.R.Schrieffer,
  Phys.Rev. {\bf 106}, 162 (1957).
  \bibitem{rice} D.Jerome, T.M.Rice, W. Kohn,
  Phys.Rev. {\bf 158}, 462 (1967).
\end{thebibliography}
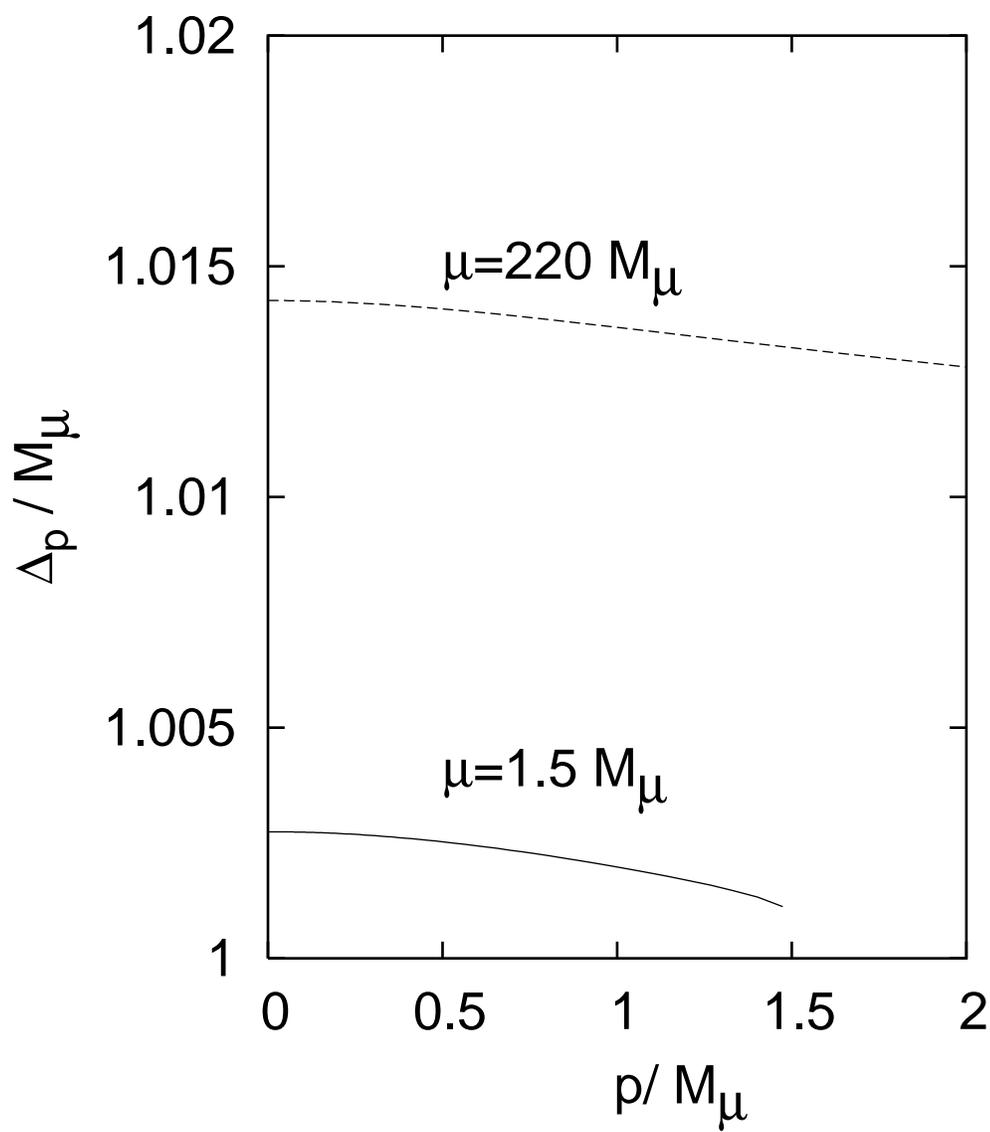
\end{document}